\documentclass[12pt]{article}
\usepackage[english,german,french,polish]{babel}
\usepackage[T1]{fontenc}
\usepackage{amsfonts}

\begin{document}

\baselineskip 0.75cm
\topmargin -0.6in
\oddsidemargin -0.1in

\let\ni=\noindent

\renewcommand{\thefootnote}{\fnsymbol{footnote}}

\newcommand{\CKM}{Cabibbo--Kobayashi--Maskawa }

\newcommand{\SM}{Standard Model }

\newcommand{\UK}{SuperKamiokande }

\pagestyle {plain}

\setcounter{page}{1}

\pagestyle{empty}

~~~

\begin{flushright}
IFT--03/1\\
hep-ph/0301161
\end{flushright}

{\large\centerline{\bf The LMA solar solution and fermion universality}}

\vspace{0.3cm}

{\centerline {\sc Wojciech Kr\'{o}likowski}}

\vspace{0.2cm}

{\centerline {\it Institute of Theoretical Physics, Warsaw University}}

{\centerline {\it Ho\.{z}a 69,~~PL--00--681 Warszawa, ~Poland}}

\vspace{0.3cm}

{\centerline{\bf Abstract}}

\vspace{0.2cm}

We suggest that the Large Mixing Angle MSW solar solution, whose unique 
physical status is confidently supported by the recent results from KamLAND 
experiments, gets its just\-ification in the fermion universality, 
interpreted for neutrinos and charged leptons in a straightforward way, most 
readily in the framework of seesaw mechanism. To this end, we consider an 
explicit seesaw model, where  Dirac and (righthanded) Majorana neutrino 
masses are simultaneously measurable, and {\it both} are conjectured to be 
proportional to charged-lepton masses. However, the LMA solar solution is 
{\it also} not inconsistent with the simple option, where neutrinos are 
Dirac particles carrying masses proportional to those of charged leptons.

\vspace{0.2cm}

\ni PACS numbers: 12.15.Ff , 14.60.Pq , 12.15.Hh .

\vspace{0.6cm}

\ni January 2003  

\vfill\eject

~~~
\pagestyle {plain}

\setcounter{page}{1}

\vspace{0.2cm}

As is known, the first results from the KamLAND long-baseline experiments 
for reactor $\bar\nu_e$'s [1] shows that the Large Mixing Angle MSW solution 
can be confidently considered [2,3,4,5,6] as the unique oscillation solution 
to the problem of solar $\nu_e$'s. The best-fit estimate is $\Delta m^2_{21} 
\equiv m^2_2 - m^2_1 \sim 7\times 10^{-5}\;{\rm eV}^2$ and $\tan^2 \theta_{12}
\sim 0.42\;(\theta_{12} \sim 33^\circ) $. The bilarge mixing matrix for 
active neutrinos $ \nu_{eL},\, \nu_{\mu L},\,\nu_{\tau L}$,

\begin{equation}
U \simeq \left(\begin{array}{ccc} c_{12} & s_{12} & 0 \\ 
-\frac{1}{\sqrt2} s_{12} & \frac{1}{\sqrt2} c_{12} & \frac{1}{\sqrt2} \\ 
\frac{1}{\sqrt2} s_{12} & -\frac{1}{\sqrt2} c_{12} & \frac{1}{\sqrt2}  
\end{array} \right)\;, 
\end{equation}

\ni where $c_{23} = 1/\sqrt2\! = s_{23}\;(\theta_{23} = 45^\circ)$ and 
$s_{13} = 0$, decribes then correctly the deficits of both solar $\nu_e$'s 
and atmospheric $\nu_\mu$'s [7] as well as the absence of oscillations of 
reactor $\bar\nu_e$'s in the Chooz experiment [8]. It gives, however, no 
LSND effect for accelerator $\bar\nu_\mu$'s and $\nu_\mu$'s [9], unless a 
third independent neutrino mass-squared scale of the order $O(1 {\rm eV}^2)$ 
is introduced beside $\Delta m^2_{21}$ and $\Delta m^2_{32}$ into the theory. 
This effect will be reinvestigated soon in the MiniBOONE experiment. The \UK 
experiments for atmospheric $\nu_\mu$'s [7] lead to the best-fit estimate 
$\Delta m^2_{32} \equiv m^2_3 - m^2_2 \sim 3\times 10^{-3}\;{\rm eV}^2$ and 
$\sin^2 2\theta_{23} \sim 1\;(\theta_{23} \sim 45^\circ )$.

In the flavor representation (used hereafter), where the charged-lepton 
mass matrix is diagonal, the mixing matrix $U$ is diagonalizing at the same 
time the active-neutrino effective mass matrix [as given in Eq. (4)].

The unique experimental status of Large Mixing Angle MSW solar solution 
requires now its theoretical explanation or, at least, its phenomenological 
correlation with other elements of neutrino physics. In this note we suggest 
that such a justification follows from the fermion universality interpreted 
for neutrinos and charged leptons in a straightforward way, where {\it both} 
Dirac and Majorana masses are involved.

To this end, let us consider an explicit model for the seesaw mechanism [10], 
where the lefthanded, Dirac and righthanded $3\times 3$ components of the 
neutrino generic $6\times 6$ mass matrix

\begin{equation}
\left( \begin{array}{cc} M^{(L)} & M^{(D)} \\ M^{(D)\,T} & M^{(R)} 
\end{array} \right) 
\end{equation}

\ni are

\begin{equation}
M^{(L)} = 0 \,,\, M^{(D)} = U\,{\rm diag}(\lambda_1,\lambda_2,\lambda_3) 
U^\dagger \,,\, M^{(R)} = -U\,{\rm diag} (\Lambda_1,\Lambda_2,\Lambda_3) 
U^\dagger\;, 
\end{equation}

\ni respectively, with $U$ as given in Eq. (1) [note that $ M^{(D)}$ and 
$ M^{(R)}$ commute {\it i.e.}, are simultaneously measurable, what 
characterizes the neutrino texture (3)]. Here, all 
$\lambda_1,\lambda_2,\lambda_3 \ll $ all $\Lambda_1,\Lambda_2,\Lambda_3$ 
and all $\geq 0$. Then, the effective $3\times 3$ mass matrices for active 
and (conventional) sterile neutrinos, $\nu_{\alpha L}$ and 
$\nu_{\alpha R} \;\;(\alpha = e,\mu,\tau)$, are

\begin{equation}
M^{(L)\,{\rm eff}} = -M^{(D)} \frac{1}{M^{(R)}} M^{(D)T} = U\;{\rm diag} 
\left( {\frac{\lambda^2_1}{\Lambda_1}}, {\frac{\lambda^2_2}{\Lambda_2}}, 
{\frac{\lambda^2_3}{\Lambda_3}} \right) U^\dagger
\end{equation}

\ni and

\begin{equation}
M^{(R)\,{\rm eff}} = M^{(R)} = -U\,{\rm diag}(\Lambda_1,\Lambda_2,\Lambda_3) 
U^\dagger \;,
\end{equation}

\ni respectively. Hence,

\begin{equation}
m_i = \frac{\lambda^2_i}{\Lambda_i} \;,\; M_i = -\Lambda_i
\end{equation}

\ni are masses (being Majorana masses) of mass neutrinos $\nu_{i L}$ and 
$\nu_{i R} \;\;(i = 1,2,3)$, respectively, connected with the flavor 
neutrinos $\nu_{\alpha L}$ and $\nu_{\alpha R} \;\;(\alpha = e,\mu,\tau)$ 
through the unitary transformations

\begin{equation}
\nu_{\alpha L}  = \sum_i U_{\alpha i}  \nu_{i L} \;,\;\nu_{\alpha R}  = 
\sum_i U_{\alpha i}  \nu_{i R} \;,
\end{equation}

\ni where $U = \left(U_{\alpha i} \right)$ gets the form (1).

The generic neutrino mass term in the Lagrangian is

\begin{equation}
- {\cal L}_{\rm mass} = \frac{1}{2} \sum_{\alpha \beta} \left( 
\overline{\nu_{\alpha L}} \,,\, \overline{(\nu_{\alpha R})^c} \right) 
\left( \begin{array}{cc} M^{(L)}_{\alpha \beta} & M^{(D)}_{\alpha \beta} \\ 
M^{(D)}_{\beta \alpha} & M^{(R)}_{\alpha \beta} \end{array} \right) 
\left( \begin{array}{c} (\nu_{\beta L})^c \\ \nu_{\beta R} \end{array} 
\right) + {\rm h.\,c.} 
\end{equation}

\ni and leads, in the case of active neutrinos, to the effective mass term

\begin{equation}
- {\cal L}^{(L)\,{\rm eff}}_{\rm mass} = \frac{1}{2} \sum_{\alpha \beta} 
\overline{\nu_{\alpha L}} M^{(L)\,{\rm eff}}_{\alpha \beta} (\nu_{\beta L})^c 
+ {\rm h.\,c.} 
\end{equation}

\ni with $M^{(L)\,{\rm eff}}$ as given in Eq. (4).

Notice that $\lambda_i$ and $\Lambda_i$ appearing in Eq. (3) are 
(simultaneously measurable) Dirac and righthanded Majorana masses, 
respectively, for the set of six mass neutrinos arising from the set of 
six flavor neutrinos which includes three active $\nu_{\alpha L}$ and three 
(conventional) sterile $\nu_{\alpha R}$. The fermion universality applied 
to neutrinos and charged leptons may mean that proportionality (at least 
approximate) occurs between their Dirac masses,

\begin{equation}
\lambda_1 {\bf :} \lambda_2 {\bf :} \lambda_3 \simeq m_e {\bf :} 
m_\mu {\bf :} m_\tau \,,
\end{equation}

\ni implying due to the first Eq. (6) the relation

\begin{equation}
m_1 {\bf :} m_2 {\bf :} m_3 \simeq \frac{m_e^2}{\Lambda_1} {\bf :} 
\frac{m_\mu^2}{\Lambda_2} {\bf :} \frac{m_\tau^2}{\Lambda_3} .
\end{equation}

\ni Here, $m_e = 0.510999$ MeV, $m_\mu = 105.658$ MeV and 
$m_\tau = 1776.99^{+0.29}_{-0.26}$ MeV [11]. Hence, 

\begin{eqnarray}
\Delta m^2_{21} & \simeq & m^2_2\left(1 - 
\frac{m_e^4 \Lambda^2_2}{m^4_\mu\Lambda^2_1}\right) =
m^2_2 \left(1 - 5.471\times 10^{-10}\,
\frac{\Lambda_2^2}{\Lambda^2_1}\right)\;, \nonumber \\
\Delta m^2_{32} & \simeq & m^2_3
\left(1 - \frac{m_\mu^4 \Lambda^2_3}{m^4_\tau\Lambda^2_2}\right) =
m^2_3 \left(1 - 1.250\times 10^{-5}\,\frac{\Lambda_3^2}{\Lambda^2_2}\right)
\end{eqnarray}

\ni and

\begin{equation}
\frac{\Delta m^2_{21}}{\Delta m^2_{32}} \simeq 
\frac{m_\mu^4 \Lambda^2_3}{m^4_\tau\Lambda^2_2}\, 
\frac{1 -m_e^4 \Lambda^2_2/m^4_\mu \Lambda^2_1}
{ 1 - m_\mu^4 \Lambda^2_3/m^4_\tau\Lambda^2_2} = 1.250\times 10^{-5}\,
\frac{\Lambda_3^2}{\Lambda^2_2}\; \frac{1 - 5.471\times 10^{-10}\,
\Lambda_2^2/\Lambda^2_1}{1 -1.250\times 10^{-5}\,
{\Lambda_3^2/\Lambda^2_2}}\,,
\end{equation}

\ni what gives

\begin{equation}
\Delta m^2_{21} \sim 3.7\times 10^{-8} \frac{\Lambda_3^2}{\Lambda^2_2}\; 
\frac{1 - 5.471\times 10^{-10}\,\Lambda_2^2/\Lambda^2_1}
{1 -1.250\times 10^{-5}\,{\Lambda_3^2/\Lambda^2_2}}\;\,{\rm eV}^2
\end{equation}

\ni with the use of the \UK estimate $\Delta m^2_{32} \sim 
3\times 10^{-3}\;\,{\rm eV}^2$.

Since $\Lambda_i$ are much larger than $\lambda_i$, it may seem that 
$\Lambda_i$ are nearly degenerate: $\Lambda_1 \simeq \Lambda_2 \simeq 
\Lambda_3$. In this case, $\Delta m^2_{21} \simeq m^2_2$ and $\Delta m^2_{32} 
\simeq m^2_3$ from Eqs. (12), while Eq. (14) predicts the value

\begin{equation}
\Delta m^2_{21} \sim 3.7\times 10^{-8}\;\,{\rm eV}^2
\end{equation}

\ni lying in the range of the LOW MSW solar solution and so, being much 
smaller than the correct Large Mixing Angle MSW value $\Delta m^2_{21} 
\sim 7\times 10^{-5}\;\,{\rm eV}^2$. Here,

\begin{equation}
m^2_1 \sim 2.0\times 10^{-17}\;\,{\rm eV}^2\;,\; m^2_2 \sim 3.7
\times 10^{-8}\;\,{\rm eV}^2 \;,\;m^2_3 \sim 3\times 10^{-3}\;\,{\rm eV}^2
\end{equation}

\ni and, when normalizing $\lambda_1 = m_e$ ({\it i.e.}, 
$\Lambda_1 = \lambda^2_1/ m_1 = m_e^2/m_1$), one gets

\begin{equation}
\Lambda_1\simeq \Lambda_2 \simeq \Lambda_3 \sim 5.8\times 10^{10}\; 
{\rm GeV}\,.
\end{equation}

In such a situation, let us conjecture tentatively that the fermion 
universality in the context of neutrinos and charged leptons has to be 
interpreted as the proportionality (at least approximate) between {\it both} 
their Dirac and Majorana [12] masses,

\begin{equation}
\lambda_1 {\bf :} \lambda_2 {\bf :} \lambda_3 \simeq \Lambda_1 {\bf :} 
\Lambda_2 {\bf :} \Lambda_3 \simeq m_e {\bf :} m_\mu {\bf :} m_\tau \,,
\end{equation}

\ni implying through the first Eq. (6) that

\begin{equation}
m_1 {\bf :} m_2 {\bf :} m_3 \simeq \frac{m_e^2}{\Lambda_1} {\bf :} 
\frac{m_\mu^2}{\Lambda_2} {\bf :} \frac{m_\tau^2}{\Lambda_3}
\simeq m_e {\bf :} m_\mu {\bf :} m_\tau .
\end{equation}

\ni In this case,

\begin{eqnarray}
\Delta m^2_{21} & \simeq & m^2_2\left(1 - \frac{m_e^2}{m^2_\mu}\right) = 
m^2_2 \left(1 - 2.339\times 10^{-5} \right)\;, \nonumber \\
\Delta m^2_{32} & \simeq & m^2_3\left(1 - \frac{m_\mu^2}{m^2_\tau}\right) = 
m^2_3 \left(1 - 3.535\times 10^{-3}\right)
\end{eqnarray}

\ni and

\begin{equation}
\frac{\Delta m^2_{21}}{\Delta m^2_{32}} \simeq 
\frac{m^2_\mu - m^2_e}{m^2_\tau - m^2_\mu} =3.548\times 10^{-3}\,,
\end{equation}

\ni predicting the value

\begin{equation}
\Delta m^2_{21} \sim 1.1\times 10^{-5}\;{\rm eV}^2\,,
\end{equation}

\ni when the \UK~estimate $\Delta m^2_{32} 
\sim 3\times 10^{-3}\;{\rm eV}^2$ is used. Here,

\begin{equation}
m^2_1 \sim 2.5\times 10^{-10}\;\,{\rm eV}^2\;,\; m^2_2 
\sim 1.1\times 10^{-5}\;\,{\rm eV}^2 \;,\;m^2_3 
\sim 3\times 10^{-3}\;\,{\rm eV}^2
\end{equation}

\ni and, when normalizing $\lambda_1 = m_e$ ({\it i.e.}, 
$\Lambda_1 = \lambda^2_1/ m_1 = m_e^2/m_1$), one obtains

\begin{equation}
\Lambda_1\sim 1.7\times 10^{7}\; {\rm GeV}\,,\, \Lambda_2 
\sim 3.4\times 10^{9}\; {\rm GeV}\,,\, \Lambda_3 
\sim 5.8\times 10^{10}\; {\rm GeV}\,.
\end{equation}

Concluding, we can see that the prediction (22) is not very 
different from the correct Large Mixing Angle MSW value 
$\Delta m^2_{21} \sim 7\times 10^{-5}\;{\rm eV}^2$. In order 
to get this value more precisely, one ought to put in Eq. (14)

\begin{equation}
\frac{\Lambda^2_3}{\Lambda^2_2} \sim 
\frac {7\times 10^{-5}\;{\rm eV}^2}{3.7\times 10^{-8}\;{\rm eV}^2} 
= 1.9\times 10^3 = 6.6 \frac{m^2_\tau}{m^2_\mu}
\end{equation}

\ni in place of the simple proportion $\Lambda^2_3/\Lambda^2_2 
\simeq m^2_\tau/m^2_\mu$, where $m^2_\tau /m^2_\mu = 282.9$. 
Then, $\Lambda_3/\Lambda_2 \sim 43 \simeq 2.6 m_\tau /m_\mu $ in 
place of $ \Lambda_3/\Lambda_2 \simeq  m_\tau /m_\mu $, where 
$m_\tau /m_\mu = 16.82 $.

Eventually, it is interesting to remark that, if neutrinos were Dirac 
particles rather than Majorana particles {\it i.e.}, $M^{\rm eff} = 
M^{(D)} = U\,{\rm diag}(\lambda_1, \lambda_2, \lambda_3) U^\dagger $ 
with $ m_i = \lambda_i $, the fermion universality for neutrinos and 
charged leptons might be interpreted as the proportionality (at least 
approximate) between their (Dirac) masses,

\begin{equation}
\lambda_1 {\bf :} \lambda_2 {\bf :} \lambda_3 \simeq m_e {\bf :} 
m_\mu {\bf :} m_\tau \,,
\end{equation}

\ni where $ m_i = \lambda_i$. Then, also in this case Eq. (21) would hold, 
predicting the value (22) for $\Delta m^2_{21} $ [13] which is not so 
different from the correct Large Mixing Angle MSW value $\Delta m^2_{21} 
\sim 7\times 10^{-5} {\rm eV}^2$. One should stress that the value (22) 
is a parameter-free prediction following from the proportionality (18) 
or (26) for Majorana or Dirac neutrinos, respectively, as they are 
investigated in the present neutrino-oscillation experiments.

Finally, let us note that writing

\begin{equation}
m_1 = \stackrel{0}{m} - \delta\;,\;m_2 = \stackrel{0}{m} + 
\delta\;,\;m_3 = \stackrel{0}{m} + \Delta 
\end{equation}

\ni and using Eq. (1) for $U$ we obtain the formula [14]

\begin{eqnarray}
M^{(L)\,\rm eff} &\!\! = \!\!& U \left( 
\begin{array}{ccc} m_{1} & 0 & 0 \\ 0 & m_{2} & 0 \\ 0 & 0 & m_3 \end{array} 
\right) U^\dagger = \,\stackrel{0}{m} 
\left( \begin{array}{ccc} 1 & 0 & 0 \\ 0 & 1 & 0 \\ 0 & 0 & 1 \end{array} 
\right) + \Delta 
\left( \begin{array}{ccc} 0 & 0 & 0 \\ 0 & \frac{1}{2} & \frac{1}{2} \\ 
0 & \frac{1}{2} & \frac{1}{2} \end{array} \right) \nonumber \\ & & \nonumber 
\\ & & \!\!+ \delta \left( \begin{array}{rrr} -\cos 2\theta_{12} & 
\frac{1}{\sqrt2} \sin 2\theta_{12} & -\frac{1}{\sqrt2} \sin 2\theta_{12} \\ 
\frac{1}{\sqrt2} \sin 2\theta_{12} & \frac{1}{2} \cos 2\theta_{12} & 
-\frac{1}{2} \cos 2\theta_{12} \\ -\frac{1}{\sqrt2} \sin 2\theta_{12} & 
-\frac{1}{2} \cos 2\theta_{12} & \frac{1}{2} \cos 2\theta_{12} \end{array} 
\right)  \;.
\end{eqnarray}

\ni In the case of prediction (22) we have from Eq. (23)

\begin{equation}
m_1 \sim 1.6\times 10^{-5}\;{\rm eV} \;,\;m_2 
\sim 3.3\times 10^{-3}\;{\rm eV} \;,\;m_3 
\sim 5.5\times 10^{-2}\;{\rm eV} \,,
\end{equation}

\ni while we get

\begin{equation}
m_2 \sim 8.4\times 10^{-3}\;{\rm eV} \;,\;m_3 
\sim 5.5\times 10^{-2}\;{\rm eV}
\end{equation}

\ni if $m^2_1 \ll m^2_2$ and we use the experimental estimate 
$\Delta m^2_{21} \sim 7\times 10^{-5}\;{\rm eV}^2$ (and also 
$\Delta m^2_{32} \sim 3\times 10^{-3}\;{\rm eV}^2$, as before). 
Due to the experimental estimate $\theta_{12} \sim 33^\circ $ we 
can put in the formula (27) $\cos 2\theta_{12} \sim 0.41$ and 
$\sin 2\theta_{12} \sim 0.91$.

\vfill\eject

~~~~
\vspace{0.5cm}

{\centerline{\bf References}}

\vspace{0.5cm}

{\everypar={\hangindent=0.6truecm}
\parindent=0pt\frenchspacing

{\everypar={\hangindent=0.6truecm}
\parindent=0pt\frenchspacing

[1]~K. Eguchi {\it et al.} (KamLAND collaboration), {\tt hep--ex/0212021}.

\vspace{0.2cm}

[2]~V. Barger and D. Marfatia, {\tt hep--ph/0212126}.

\vspace{0.2cm}

[3]~G.L. Fogli {\it et al.}, {\tt hep--ph/0212127}.

\vspace{0.2cm}

[4]~M. Maltoni, T. Schwetz and J.W.F. Valle, {\tt hep--ph/0212129}.

\vspace{0.2cm}

[5]~A. Bandyopadhyay {\it et al.}, {\tt hep--ph/0212146v2}.

\vspace{0.2cm}

[6]~J.N.~Bahcall, M.C.~Gonzalez--Garcia and C. Pe\~{n}a--Garay, 
{\tt hep--ph/0212147v2}.

\vspace{0.2cm}

[7]~S. Fukuda {\it et al.}, {\it Phys. Rev. Lett.} {\bf 85}, 3999 (2000).

\vspace{0.2cm}

[8]~M. Appolonio {\it et al.}, {\it Phys. Lett.} {\bf B 420}, 397 (1998); 
{\bf B 466}, 415 (1999).

\vspace{0.2cm}

[9]~G. Mills, {\it Nucl. Phys. Proc. Suppl.} {\bf 91}, 198 (2001); 
{\it cf.} also K. Eitel, {\it Nucl. Phys. Proc. Suppl.} {\bf 91}, 191 (2001) 
for negative results of KARMEN2 experiment.

\vspace{0.2cm}

[10]~M. Gell-Mann, P. Ramond and R.~Slansky, in  {\it Supergravity}, edited 
by F.~van Nieuwenhuizen and D.~Freedman, North Holland, 1979; T.~Yanagida, 
Proc. of the {\it Workshop on Unified Theory and the Baryon Number in the 
Universe}, KEK, Japan, 1979; R.N.~Mohapatra and G.~Senjanovi\'{c}, 
{\it Phys. Rev. Lett.} {\bf 44}, 912 (1980).

\vspace{0.2cm}

[11]~The Particle Data Group, {\it Phys.~Rev} {\bf D 66}, 010001 (2002).

\vspace{0.2cm}

[12]~The possibility that the eigenvalues of $M^{(R)}$ are hierarchical 
was considered previously, {\it cf.} S.F.~King {\it Nucl. Phys.} 
{\bf B 562}, 57 (1999); {\bf B 576}, 85 (2000); {\tt hep--ph/0208266}; 
{\it cf.} also G.~Altarelli, F.~Feruglio and I. Masina, 
{\tt hep--ph/0210342}, and references therein.

\vspace{0.2cm}

[13]~W. Kr\'{o}likowski, {\tt hep--ph/0210417}.

\vspace{0.2cm}

[14]~W. Kr\'{o}likowski, {\tt hep--ph/0208210}.

\vfill\eject

\end{document}